\documentclass[12pt,a4paper,final]{iopart}

\usepackage{iopams}
\usepackage{graphicx,cite}
\usepackage[breaklinks=true,colorlinks=true,linkcolor=blue,urlcolor=blue,citecolor=blue]{hyperref}

\begin{document}
	
	\title[Floquet geometric entangling gates in ground-state manifolds of Rydberg atoms]{Floquet geometric entangling gates in ground-state manifolds of Rydberg atoms}
	
	\author{Hao-Wen~Sun, Jin-Lei~Wu$^{\ast}$ and Shi-Lei~Su$^{\ast}$}
	\address{School of Physics, Zhengzhou University, Zhengzhou 450001, China}
	
	\address{$\ast$~Author to whom any correspondence should be addressed.}
	\ead{jlwu517@zzu.edu.cn}\ead{slsu@zzu.edu.cn}
	
\begin{abstract}
We propose new applications of Floquet theory in Rydberg atoms for constructing quantum entangling gates in atomic ground-state manifolds. By dynamically periodically modulating the Rabi frequencies of transitions between ground and Rydberg states of atoms, error-resilient two-qubit entangling gates can be implemented in the regime of Rydberg blockade. According to different degrees of Floquet theory utilization, the fidelity of the resulting controlled gates surpasses that of the original reference, and it exhibits high robustness against Rabi error in two qubits and detuning error in the control qubit. Our method only uses encoding in the ground states, and compared to the original 
scheme using Rydberg state for encoding, it is less susceptible to environmental interference, making it more practical to implement. Therefore, our approach may have broader applications or potential for further expansion of geometric quantum computation with neutral atoms.
	\end{abstract}
	
	%\pacs{}
	\vspace{2pc}
	%\submitto{}
	%\maketitle
	\section{Introduction}
As one of the main themes in physics of this century, the geometric phase~\cite{berry1984quantal,aharonov1987phase,wilczek1984appearance,anandan1988non} has received much attention in quantum computing. Quantum computing can provide more efficient solutions than traditional computers in areas such as prime factorization~\cite{vandersypen2001experimental,xu2012quantum,martin2012experimental}
and machine learning~\cite{rebentrost2014quantum,li2015experimental,cong2019quantum}.
Because a geometric phase depends only on the evolution path, quantum gates constructed through geometric phases have better anti-internal noise interference ability~\cite{de2003berry,zhu2005geometric,leek2007observation,filipp2009experimental,berger2013exploring}. The geometric phase theory of quantum systems is the foundation of various patterns of geometric quantum computation. Starting from the seminal adiabatic geometric quantum computation~(GQC)~\cite{jones2000geometric,wu2005holonomic,wu2013geometric,huang2019experimental} based on Berry phases and adiabatic holonomic quantum computation~\cite{zanardi1999holonomic,duan2001geometric,Wu_2019} based on adiabatic non-Abelian geometric phases, it has gradually developed to include nonadiabatic geometric quantum computation~(NGQC)~\cite{xiang2001nonadiabatic,zhu2002implementation,thomas2011robustness,
	zhao2017rydberg,li2020approach,chen2018nonadiabatic,zhang2020high} based on nonadiabatic Abelian geometric phases, and non-adiabatic holonomic geometric quantum computation~(NHQC)~\cite{liu2019plug,sjoqvist2012non,xu2012nonadiabatic,xue2015universal,xue2017nonadiabatic,zhou2018fast,hong2018implementing,mousolou2017electric,zhao2020general,johansson2012robustness,zheng2016comparison,ramberg2019environment,jing2017non,liu2021super} based on nonadiabatic non-Abelian geometric 
phases. Due to the limited ability of these non-adiabatic gates 
to address laser control errors, to tackle this issue, the application of periodic control pulses in the 
GQC scenario is gradually increasing~\cite{novivcenko2017floquet,novivcenko2019non,bomantara2018quantum,bomantara2018simulation}.

Recently, a new calculation scheme, Floquet geometric quantum computation~(FGQC)~\cite{wang2021error}, has gained significant attention, where universal 
error-resistant geometric gates can be constructed via a non-Abelian geometric phase. In the framework of traditional GQC scheme, add a periodic function that rapidly changes over time to the Rabi frequency $\Omega_0$. In this scenario, the effective Hamiltonian of the original Hamiltonian is determined, 
simplified, and reduced to a form suitable for straightforward calculation. This effective Hamiltonian enables the determination of parameters crucial for constructing the 
corresponding geometric entangling gates. The theoretical framework for solving these gate parameters is referred to as Floquet theory. FGQC has recently 
been demonstrated in experiments with utracold atoms~\cite{cooke2024investigation}. However, although the original Floquet two-qubit scheme using atomic Rydberg states for encoding qubits can achieve gate operations in a single step, 
Rydberg states are necessarily manipulated in low-temperature and high-vacuum environments, which 
means they are highly sensitive to external interference and may affect the accuracy of computational results. Furthermore, it is evident from the matrix expression given in equation~(\ref{eq.B8}) that the original Floquet scheme cannot imple ructing controlled-X gate and controlled-T gate. We discard the approach of using excited state encoding in favor of ground state encoding, thereby applying the Floquet theory. 
And if two of atomic ground states are used for encoding qubit, they are easier to 
prepare and control than Rydberg states, making the gate implementation simpler, less sensitive 
to external interference, more feasible in practical applications.

In this paper, we apply Floquet theory to construct two-qubit geometric entangling gates in ground-state manifolds of Rydberg atoms in the regime of Rydberg blockade~\cite{jaksch2000fast,saffman2010quantum,petrosyan2017high,levine2018high,PhysRevApplied.16.064031}. The  performances of resulting quantum gates are analyzed with respect 
to different degrees of Floquet engineering, decay rate, and different 
quantum gates. We compare the proposed Floquet gate scheme with the original one, and perform numerical simulations in the presence of certain system control errors.
The numerical results indicate that, compared with the reference Rydberg-state Floquet gate, the present ground-state Floquet gate can gain higher fidelity and better robustness 
to control errors. When ignoring the atomic decay, using the Floquet scheme for both the control and target qubits will achieve higher fidelity and robustness. 
When there is a decay rate, using the Floquet scheme only for the target qubit will achieve higher fidelity while maintaining high robustness. In addition, the controlled phase gate has high robustness to decay rates, which has been analyzed and demonstrated in the article.

This paper is organized as follows: In Sec.~\ref{sec:2}, we analyzed the general theory of Floquet geometric entangling gates in the regime of Rydberg blockade. 
In Sec.~\ref{sec:3}, we introduced and derived two new schemes with the Floquet scheme in Rydberg-blockade gates, demonstrating the feasibility of the proposed schemes. In Sec.~\ref{sec:4}, we select two sets of possible parameters, providing numerical simulations for both corresponding schemes to validate their performance. We examine the robustness with respect to Rabi error $\delta$ and the fidelity in different schemes for CNOT and CT gates. Subsequently, 
we analyze the impact of spontaneous emission on the gate schemes, identify the relatively superior scheme in each scenario, and offer optimization suggestions. The conclusion is presented in Sec.~\ref{sec:5}.
\section{Physical model with two Rydberg atoms}\label{sec:2}
% \begin{figure}
% 	\centering%居中
% 	\includegraphics[width=0.4\textwidth]{figure1.png}
% 	\caption{Energy-level diagram of a three-level, two-qubit Rydberg atom system. 
% 	The $|r\rangle$ state corresponds to a Rydberg state, while the
% 	 $|g\rangle$ state and $|e\rangle$ state denote the ground states. 
% 	 The control qubit of this three-level system is driven by a square wave $\Omega_1$ in the DG-FGQC 
% 	 scheme and by a phase-modulated periodic pulsed $f_1(t)\Omega_1e^{-i\varphi_c}$ in the FGQC-FGQC scheme, 
% 	 but ultimately both result in an effective $\pm\pi$ phase. 
% 	 The target qubit is driven by periodic pulses $\Omega_{0r}(t)e^{-i\varphi_0(t)}$ and $\Omega_{1r}(t)e^{-i\varphi_1(t)}$ 
% 	 with time-dependent phase, where V represents the RRI intensity. The effect of the interaction is shown 
% 	 in the figure.}%图片标题
% 	\label{fig:1}%图片标识
% \end{figure}
	\begin{figure}[t]\centering
	\includegraphics[width=0.7\linewidth]{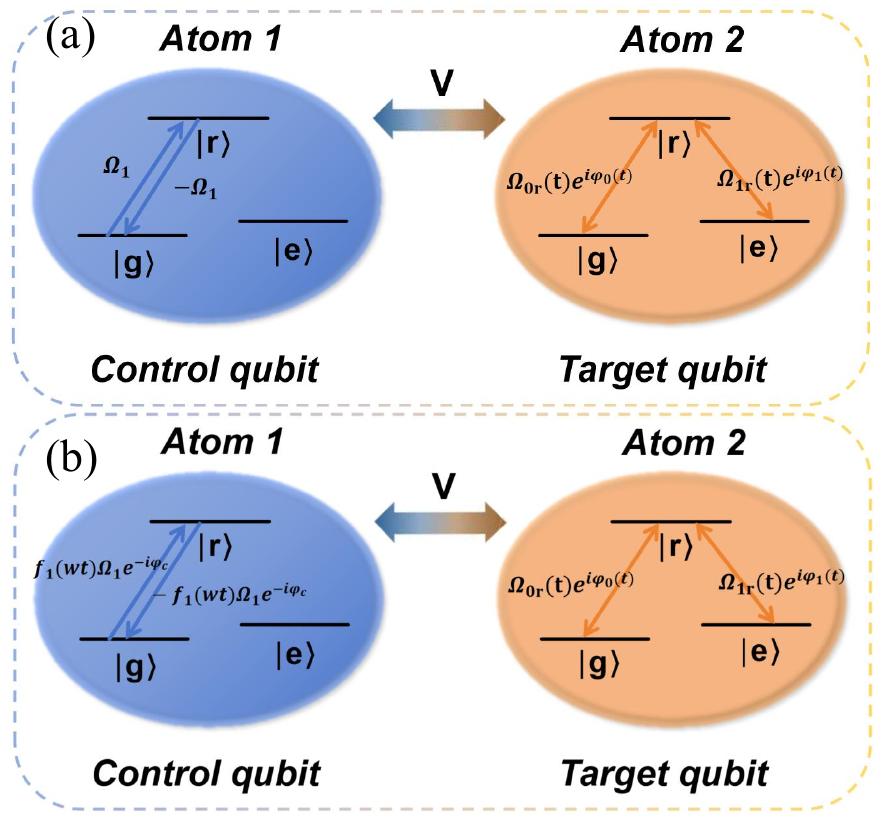}\\ 
	\caption{Energy level diagram of two Rydberg atoms with RRI strength $V$ for constructing the two-qubit entangling gates in (a)~scheme A and (b)~scheme B.}\label{fig1}
\end{figure}
Consider a two-atom system where two atoms interact with each other through the van der Waals Rydberg-Rydberg interaction~(RRI). For constricting a conditioned two-qubit 
entangling gate, we specify the control and target qubits, as shown in figures~\ref{fig1}(a) and \ref{fig1}(b). The two atoms have two ground states $|g\rangle$ and $|e\rangle$ 
encoding qubits and a Rydberg state $|r\rangle$ mediating RRI between the permanent dipole moments of Rydberg states excited by a constant electric field, which are far greater 
than the laser Rabi frequency, to entangle atoms. 

To implement gate operations under distributed laser control, we conduct a three-step scheme in the regime of Rydberg blockade. (i) Apply a $\pi$-pulse on the control qubit 
to induce the atomic transition from the ground $|g\rangle$ to the Rydberg state $|r\rangle$, with Rabi frequency $\Omega_1$. (ii) Apply pulses on the target qubit that 
drives separately the transitions from $|g\rangle$ and $|e\rangle$ states to the Rydberg state, with time-dependent Rabi frequencies $\Omega_{0r}$(t) and $\Omega_{1r}$(t), 
respectively. (iii) Apply a $-\pi$-pulse on the control qubit to deexcite the atom from the Rydberg state to the ground state $|g\rangle$, with a Rabi frequency $-\Omega_1$. 
For the system remaining in the Rydberg blockade regime, a strong RRI strength is assumed $V\gg\Omega_1$, $\Omega_{0r}(t)$, $\Omega_{1r}(t)$. In the Rydberg blockade regime, 
the system Hamiltonian containing the three steps can be written as
\begin{eqnarray}\label{eq.1}
H(t) &=& \frac{\Omega_1}{2} | g \rangle_1 \langle r |\otimes \mathbb{I} + \frac{\Omega_{0r}(t)}{2} e^{i\varphi_0(t)}| e \rangle_1 \langle e |
\otimes | g \rangle_2 \langle r |\nonumber\\ 
&&+ \frac{\Omega_{1r}(t)}{2} e^{i\varphi_1(t)}| e \rangle_1 \langle e |
\otimes | e \rangle_2 \langle r |
+ {\rm H.c.}
\end{eqnarray}
$\mathbb{I}$ is the identity operator acting on the Rydberg atom.

To achieve a Floquet two-qubit gate, we introduce a periodic 
function $f(wt)=\cos(wt)$ and a constant $\theta$, and set $\Omega_{0r}(t)=\sin(\theta/2) f(wt)\Omega_0$ and 
$\Omega_{1r}(t)=\cos(\theta/2) f(wt)\Omega_0$. 
Then the  laser pulses of target qubit can be written as  
$\Omega(t)=\sqrt{ {\Omega}^{2}_{0r}(t)+{\Omega}^{2}_{1r}(t) }=f(wt)\Omega_0$, 
where $\tan(\theta/2) = \Omega_{0r}$(t)/ $\Omega_{1r}$(t). Expressing $| g \rangle \otimes | e \rangle$ 
in a reduced form $|ge \rangle$, then rewrite the Hamiltonian in 
Eq. (\ref{eq.1}) as
\begin{eqnarray}\label{eq.2}
H(t)&=\frac{\Omega_1}{2} | g \rangle_1 \langle r | \otimes \mathbb{I}\nonumber\\ 
     &+\frac{f(wt)}{2} \Omega_0\left [ \sin\frac{\theta}{2} e^{i\varphi_0(t)}| eg \rangle \langle er |+ \cos\frac{\theta}{2} e^{i\varphi_1(t)} | ee \rangle \langle er |\right ] 
     + {\rm H.c.}		
\end{eqnarray}
Based on this Hamiltonian, we next show two different schemes to implement Floquet geometric entangling gates in the Rydberg blockade regime.

\section{Floquet geometric entangling gates}\label{sec:3}
In former scheme of FGQC~\cite{wang2021error}, the qubit contain a Rydberg state, for which such an encoding way makes it easier to achieve coupling between qubits but makes the quantum gates more susceptible to environmental interference compared to the gates with ground-state qubits, because Rydberg states would suffer higher decoherence rates and have a relatively shorter lifetime. Moreover, It requires precise experimental conditions and a high degree of control accuracy to manipulate Rydberg states, demanding advanced experimental equipment and techniques. Therefore, here we utilize two ground states to encode qubit states as $|0\rangle\equiv|g\rangle$ and $|1\rangle\equiv|e\rangle$ 
respectively, with the state $|r\rangle$ serving solely as an intermediate state rather than a computational state. Simultaneously providing $\mathbb{I} = | 0 \rangle \langle 0 |+| 1 \rangle \langle 1 |$, 
then we can obtain
\begin{eqnarray}\label{eq.3}
H(t)&=& \frac{\Omega_1}{2} ( | 00 \rangle \langle r0 |
+| 01 \rangle \langle r1 |)+ \frac{f(wt)}{2} \Omega_0\left[ \sin\frac{\theta}{2} e^{i\varphi_0(t)}| 10 \rangle \langle 1r |\right.\nonumber\\  
&&+ \left.\cos\frac{\theta}{2} e^{i\varphi_1(t)} | 11 \rangle \langle 1r | \right] + {\rm H.c.}		
\end{eqnarray}
\subsection{Scheme A: Target-qubit Floquet engineering}\label{subsec:3A}
In this scheme, we combine the standard dynamical gate (DG) with FGQC, referred to DG-FGQC. In steps (i) and (iii), we employ the DG method, while in step (ii), we utilize Floquet theory. We initially apply the $\pi$ pulses $\Omega_1$, resonant with the transition frequency between $| 0 \rangle$ and $|r \rangle$, in a square wave form on the control 
qubit. The state $| 1 \rangle$ is not affected by the laser. If the control qubit 
is initially in the state $| 0 \rangle$, the pulse sequence $\Omega_1$ will bring the control qubit to the Rydberg state, while no change occurs if the initial state of the control qubit is $| 1 \rangle$. The pulse duration is $T_1=\pi / \Omega_1$. Then moving on to the target qubit, due to the presence of RRI, within the Rydberg blockade regime $V\gg\Omega_0$, the target qubit cannot be excited to the 
Rydberg state when the control qubit is initially in the state $| 0 \rangle$. When the control qubit is initially in the state $| 1 \rangle$, the Hamiltonian on the target state in the second step is
\begin{equation}\label{eq.4}
H_t(t) =\frac{f(wt)}{2} \Omega_0e^{i\varphi_1(t)}\left[ \sin\frac{\theta}{2} e^{i\varphi(t)}|0 \rangle
+ \cos\frac{\theta}{2}  |1 \rangle\right]\langle r | + {\rm H.c.},
\end{equation}
with $\varphi_0(t)=\varphi_1(t)+\varphi(t)$. Then we can define a bright state $| b \rangle=\sin(\theta/2) e^{i\varphi(t)}| 0 \rangle
+ \cos(\theta/2)  | 1 \rangle$, and a decoupled dark state $| d \rangle= \cos(\theta/2) e^{i\varphi(t)}| 0 \rangle- \sin(\theta/2)  | 1 \rangle$~\cite{sun2021one}. 
Obviously, the effective Hamiltonian 
only affects the subspace $S=Span \left \{   | b \rangle, | r \rangle\right \}$, and then the Hamiltonian~(\ref{eq.4}) can be rewritten as follows
\begin{equation}\label{eq.5}
H_t(t) = f(wt)\mathbf{F}\cdot \mathbf{n}(t),
\end{equation}
where $\mathbf{F}\equiv \left(\tilde{\sigma}_x,\tilde{\sigma}_y,\tilde{\sigma}_z\right)^T/2$ being the spin operators on the subspace $S$ 
with $\tilde{\sigma}_x=\left | b  \right \rangle \left\langle r\right| +\left | r  \right \rangle \left\langle b\right|$, 
$\tilde{\sigma}_y=-i\left | b  \right \rangle \left\langle r\right| +i\left | r  \right \rangle \left\langle b\right|$, 
$\tilde{\sigma}_z=\left | b  \right \rangle \left\langle b\right| -\left | r  \right \rangle \left\langle r\right|$, 
and $\mathbf{n}(t)\equiv\Omega_0\left(\cos\varphi_1(t),-\sin\varphi_1(t),0\right)^T$ with the unit vector $\left(\cos\varphi_1(t),-\sin\varphi_1(t),0\right)^T$. 

According to reference~\cite{wang2021error}, 
we select 
$R(t)=\exp[-i\sin(wt)\mathbf{F}\cdot \mathbf{n}(t)/w]$, 
which satisfies the boundary condition $R(\tau) =  R(0) = \mathbb{I}$ if $\omega \tau = k\pi, \forall k \in \mathbb{N}^+$. 
Compared to \ref{eqA}, we set $\lambda_1(t)=wt$, $\lambda_2(t)=\varphi(t)$, $H_0(\lambda_2)=\mathbf{F}\cdot \mathbf{n}(\lambda_2)$ and 
$\mathcal{F}(\lambda_1)=\frac{\sin(wt)}{w} $ where $\mathcal{F}(\lambda_1)$ is the original function of $f(\lambda_1)=\cos(wt)$. So we obtain a pure geometric effective Hamiltonian
\begin{equation}\label{eq.6}
H_t(t) =  \dot{\varphi}(t)R^{\dagger}(t)(-i\partial /\partial\varphi)R(t),
\end{equation}
with $R(t)=\exp[-i\mathcal{F}(\lambda_1)H_0(\lambda_2)]$. The subsection C in section \uppercase\expandafter{\romannumeral2} in reference~\cite{wang2021error} provides a brief introduction to 
the work reported in reference~\cite{novivcenko2019non}, which indicates that in a periodically driven system with 
$H(t)=f[\lambda_1(t)]H_0[\lambda_2(t)]$, the adiabatic evolution of quantum systems within completely degenerate Floquet bands will 
result in the formation of non-Abelian geometric phases. Following the method in reference~\cite{novivcenko2019non},
when $w\gg\dot{\varphi}(t)$, and the Hamiltonian varies rapidly with $\lambda_1$ while changing slowly with 
$\lambda_2$, so the adiabatic operator $\lambda_2$ does not depend on $\lambda_1$. We solve equation~(\ref{eq.6}) for the partial 
derivative with respect to $\mathcal{F}(\lambda_1)$. 
\begin{equation}\label{eq.20}
	\frac{\partial H_t(\mathcal{F},t)}{\partial \mathcal{F}(\lambda_1)} =H_0R^{\dagger}\frac{\partial}{\partial t} R-R^{\dagger}\frac{\partial}{\partial t}(RH_0). 
\end{equation}
Since $R^{\dagger}\frac{\partial}{\partial t}(RH_0)=\dot{H_0}+iH_{t}H_0$, then we obtain
\begin{equation}\label{eq.21}
	\frac{\partial H_t(\mathcal{F},t)}{\partial \mathcal{F}(\lambda_1)} =-\dot{H_0}+i[H_0,H_t]. 
\end{equation}

To calculate the effective Hamiltonian $H_{eff}(t)$ more conveniently, following the form $H_0(t) = \mathbf{F} \cdot \mathbf{n}(t)$, 
we set $H_t(\mathcal{F},t) = \mathbf{F} \cdot \mathbf{X}$ and substitute it into equation~(\ref{eq.21}) and subtract $\mathbf{F}$ from both sides of the equation to obtain  
\begin{equation}\label{eq.22}
	\frac{\partial \mathbf{X}}{\partial \mathcal{F}} =-\dot{\mathbf{n}}+\mathbf{n}\times \mathbf{X}, 
\end{equation}
with the initial condition $\mathbf{X}(\mathcal{F}, t) = 0$ for $\mathcal{F} = 0$. A solution to this equation is
\begin{equation}\label{eq.23}
\mathbf{X}(\mathcal{F},t)=-\mathcal{F}\frac{(\mathbf{n} \cdot\dot{\mathbf{n}})\mathbf{n}}{n^2}
-\sin(\mathcal{F}n)\frac{(\mathbf{n} \times{\dot{\mathbf{n}}})\times \mathbf{n}}{n^3}
-[\cos(\mathcal{F}n)-1]\frac{(\mathbf{n} \times{\dot{\mathbf{n}}})}{n^2}, 
\end{equation}
where $n=\left |\mathbf{n}\right |$. Then, we can obtain Hamiltonian through $H_t(\mathcal{F},t) = F \cdot X$ and $\mathcal{F}(wt)=\frac{\sin(wt)}{w} $:
\begin{eqnarray}\label{eq.24}
H_t(t)=&-\frac{\sin(wt)}{w} \frac{(\mathbf{n} \cdot \dot{\mathbf{n}})\mathbf{n}\cdot \mathbf{F}}{n^2}\nonumber\\
&-\sin\left [\sin(wt)\frac{n}{w}\right ]\frac{(\mathbf{n} \times{\dot{\mathbf{n}}})\times \mathbf{n}\cdot \mathbf{F}}{n^3}\nonumber\\
&-\left\{\cos\left [\sin(wt)\frac{n}{w}\right ]-1\right\} \frac{(\mathbf{n} \times{\dot{\mathbf{n}}})\cdot \mathbf{F}}{n^2}
\end{eqnarray}
Clearly, under the periodic driving of $f(wt)=\cos(wt)$, the average of the first and second terms in equation (\ref{eq.24}) is zero, and does not contribute to the effective Hamiltonian $H_{eff}$. 
The third term of the equation contributes to the effective Hamiltonian:
\begin{equation}\label{eq.25}
	H_{\rm eff}(t)=[1-J_{0}(\frac{n}{w})]\frac{(\mathbf{n} \times{\dot{\mathbf{n}}})\cdot \mathbf{F}}{n^2}, 
\end{equation}
where $J_{0}(a)=\frac{1}{2\pi}\int_{0}^{2\pi} e^{ia\sin\theta }d\theta$ is the zero-order Bessel function. Due to 
$\mathbf{n}(t)=\Omega_0\left(\cos\varphi_1(t),-\sin\varphi_1(t),0\right)^T$, we have $n=\left | \mathbf{n} \right | =\Omega_0$, 
then the effective Hamiltonian can be written as
\begin{equation}\label{eq.7}
H_{\rm eff}(t)=[1-J_{0}(\Omega_0/w)]\mathbf{F}\cdot[\mathbf{n}(t)\times \dot{\mathbf{n}}(t)]/\Omega_0^2,
\end{equation}
Set $\varphi_1(t)=Nt$,  and then from the evolution operator $U=\mathcal{T}e^{-i\int_{0}^{\tau} H_{\rm eff}(t)dt} $ at time $t=\tau$, 
we get
\begin{eqnarray}\label{eq.8}
U(\tau,0)=\left(
\begin{array}{cc}
e^{-iC\tau}                &0\\
0                 &e^{iC\tau}\\
\end{array}
\right).
\end{eqnarray}
with $C\equiv$$-\left(N/2\right)\left [1-J_{0}(\Omega_0/w)  \right ] $, at the basis $\left \{| b \rangle, |r \rangle\right \}$ of the target qubit. 

The dark state $| d \rangle$ is constructed to be orthogonal to the bright state, and it does not change after the laser changes the bright state. So we write the evolution operator as
\begin{equation}\label{eq.9} 
U(\tau,0)=e^{-iC\tau}\left | b  \right \rangle \langle b|+
e^{iC\tau}\left | r  \right \rangle \langle r|+
\left | d  \right \rangle \langle d|.
\end{equation}
Here, since $\left |r  \right \rangle$ serves as an auxiliary intermediate state uninvolved in the qubit states, we can rewrite Eq.~(\ref{eq.9}) as
\begin{equation}\label{eq.10} 
U(\tau,0)=e^{-iC\tau}\left | b  \right \rangle \langle b|+
\left | d  \right \rangle \langle d|.
\end{equation}
To obtain the control parameters for the target qubit, we transform the two-qubit basis vectors into 
$\left \{ \left | 10  \right \rangle ,\left | 11  \right \rangle  \right \} $ and then obtain a new evolution matrix
\begin{eqnarray}\label{eq.11}
U(\tau,0)=\left(
\begin{array}{cc}
e^{-iC\tau}\sin\left(\frac{\theta }{2} \right)^2+\cos\left(\frac{\theta }{2} \right)^2                 &e^{-iC\tau}\sin\theta \frac{e^{i\varphi}}{2}-\sin\theta \frac{e^{i\varphi}}{2}\\
e^{-iC\tau}\sin\theta \frac{e^{-i\varphi}}{2}-\sin\theta \frac{e^{-i\varphi}}{2}                 &e^{-iC\tau}\cos\left(\frac{\theta }{2} \right)^2+\sin\left(\frac{\theta }{2} \right)^2\\
\end{array}
\right).
\end{eqnarray}
Then, we can choose suitable parameters to obtain the corresponding quantum gate for the target qubit.
The laser action time for this process is $T_2$. Finally, perform another $-\Omega_1$ laser on the control bit for $T_3=\pi/\Omega_1$, which returns the control qubit to its initial state. When the control qubit is in the state $\left | 0  \right \rangle$, it will be excited to the Rydberg 
state after the first step, preventing the target qubit from reaching the excited state, which leads to the blocking of target qubit, rendering the gate $U(\tau,0)$ ineffective. Overall, a conditioned two-qubit geometric gate is constructed with form $U_{CU}=|0\rangle_1\langle0|\otimes\mathbb{I}+|1\rangle_1\langle1|\otimes U(\tau,0)$.

\subsection{Scheme B: Two qubit Floquet engineering}\label{subsec:3B}
In this subsection, to simultaneously conduct Floquet engineering on the control and target qubits, we will augment the single-qubit FGQC scheme on basis of subsection \ref{subsec:3A} by incorporating the control qubit, which is referred to as FGQC-FGQC.

As the laser affecting the control qubit only influences the distribution of states 
$\left | g  \right \rangle$ and $\left | r  \right \rangle$ in the first step, we reduce the single-qubit
three-level system to a two-level system with the Hamiltonian 
\begin{equation}\label{eq.12}
H_{\rm ctrl}=\frac{\Omega_1}{2} | g \rangle \langle r | + {\rm H.c.} 
\end{equation} 
for the control atom. To construct a single-qubit Floquet gate that transforms the control qubit from the ground state 
$\left | g  \right \rangle$ to the excited state $\left | r  \right \rangle$ and makes the system return to the initial state after 
a Floquet operation on the target qubit, we treat the $\left | r  \right \rangle$ state as the computational state for 
analysis. We re-encode it, representing the $\left | g \right \rangle$ and $\left | r  \right \rangle$ state as 
$\left | 0  \right \rangle$ and $\left | 1  \right \rangle$ states, respectively. Then we set the Floquet Hamiltonian 
\begin{eqnarray}\label{eq.13}
H_{\rm ctrl}'&=&f_1\frac{\Omega_1}{2}e^{-i\varphi_{c}} | 0 \rangle \langle 1 | + {\rm H.c.}\nonumber\\
&=&f_1\frac{\Omega_1}{2}\left(\cos\varphi_c\sigma_x+\sin\varphi_c\sigma_y\right),	   	 
\end{eqnarray}
with $f_1=\cos(\lambda_1')$ and $\lambda_1'=w_1t$. In order to enable the construction of an X gate using Floquet, 
it needs to introduce a detuning term on the control qubit~\cite{wang2021error}.
Then the Hamiltonian with detuning can be rewritten as
\begin{eqnarray}\label{eq.14}
H_{f{\rm ctrl}}&=&f_1\frac{\Delta_1(t)}{2}\sigma_z+f_1\frac{\Omega_1(t)}{2}\left(\cos\varphi_c\sigma_x+\sin\varphi_c\sigma_y\right)\nonumber\\
&=&f_1\mathbf{F_0}\cdot \mathbf{r}(t),   	 
\end{eqnarray}
with $\mathbf{F_0}=\left(\sigma_x,\sigma_y,\sigma_z\right)^T/2$ and the unit vector
$\mathbf{r}=\left(\Omega_1(t)\cos\varphi_c,\Omega_1(t)\sin\varphi_c,\Delta_1(t)\right)$. $\Delta_1(t)$ and $\Omega_1(t)$ are time-dependent parameters, and we set $\Omega'=\sqrt{\Omega_1^2(t)+\Delta_1^2(t)}$. 
The variable in $\mathbf{F_0}\cdot \mathbf{r}$ that evolves slowly over time compared to $f_1$ is $\lambda_2'(t)$, and similarly, when $w_1\gg\dot{\lambda_2'}(t)$, we obtain a Floquet effective Hamiltonian
\begin{equation}\label{eq.15}
H_{c{\rm eff}}= [1-J_{0}(\Omega'/w_1)]\mathbf{F_0}\cdot[\mathbf{r}(t)\times \dot{\mathbf{r}}(t)]/\Omega'^2, 					  
\end{equation}
and then we can obtain the parameters of the X gate through the evolution operator 
$U\approx\mathcal{T}e^{-i\int_{0}^{\tau_1} H_{c{\rm eff}}(t)dt} $. Because equation~(\ref{eq.15}) have the same form as equation~(\ref{eq.7}), it can induce a single-qubit operation of matrix equation~(\ref{eq.11}), making the atomic transition from $|g\rangle$ to $|r\rangle$, and vice versa. Combined with the second-step operation on the target qubit described in the scheme A, one can implement the FGQC-FGQC conditioned two-qubit entangling gate $U_{CU}=|0\rangle_1\langle0|\otimes\mathbb{I}+|1\rangle_1\langle1|\otimes U(\tau,0)$.
\begin{figure}[htp]\centering
	\includegraphics[width=0.6\linewidth]{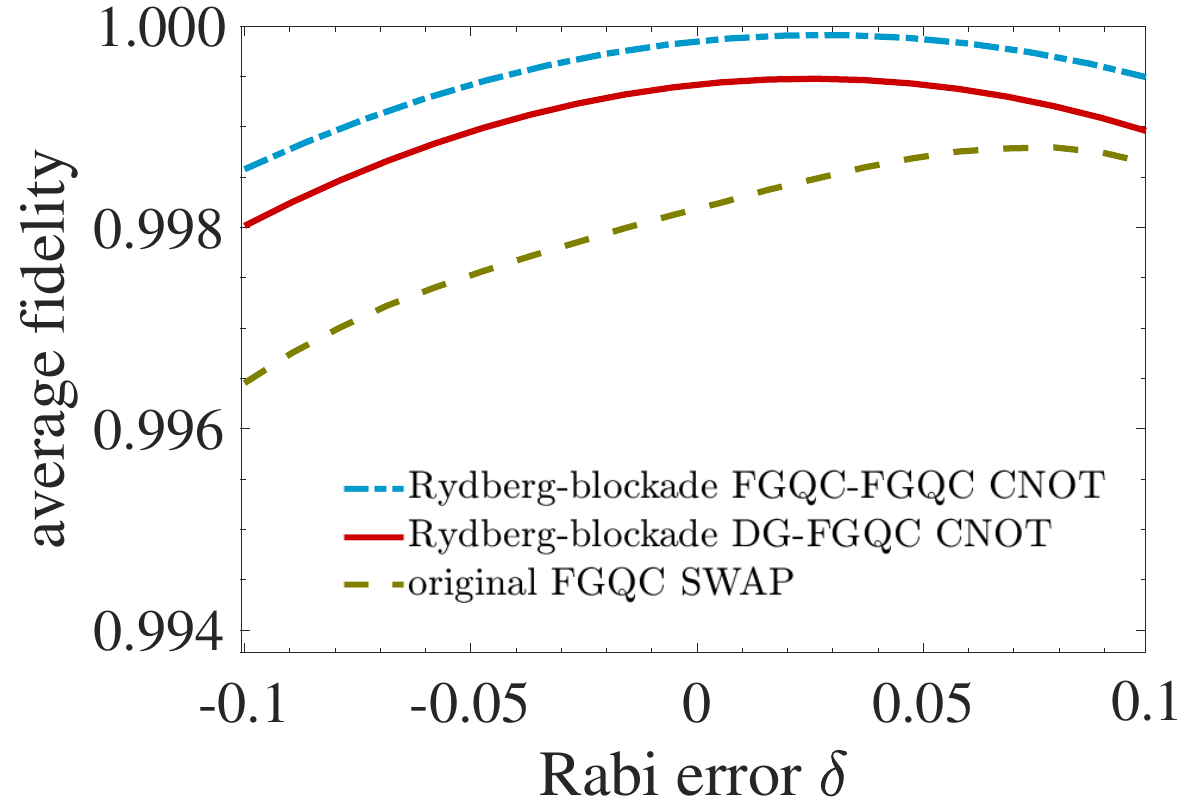}\\ 
	\caption{Comparison of the average fidelity for implementing a CNOT gate under our two new Floquet schemes, and 
	implementing a SWAP-like gate under original Floquet scheme.}\label{fig2}
\end{figure}

\section{Numerical simulations}\label{sec:4}	
In our approach, to obtain the desired two-qubit quantum gate, mainly to implement transformation in equation~(\ref{eq.11}), we set $\varphi=0$ to obtain
\begin{eqnarray}\label{eq.17}
U(\tau,0)=\left(
\begin{array}{cc}
e^{-iC\tau}\sin\left(\frac{\theta }{2} \right)^2+\cos\left(\frac{\theta }{2} \right)^2                 &e^{-iC\tau} \frac{\sin\theta}{2}- \frac{\sin\theta}{2}\\
e^{-iC\tau} \frac{\sin\theta}{2}- \frac{\sin\theta}{2}                 &e^{-iC\tau}\cos\left(\frac{\theta }{2} \right)^2+\sin\left(\frac{\theta }{2} \right)^2\\
\end{array}
\right).
\end{eqnarray}
For example, we next conduct numerical simulations of implementing CNOT gates and CT gates. For the numerical simulations in the following, the Hamiltonians for the three steps are, respectively
\begin{eqnarray}\label{three_step}
H_1&=&\mathcal{F}\left[\frac{\Omega_1}{2} |g\rangle_1 \langle r | + {\rm H.c.}+\frac{\Delta_1}{2}(|g\rangle_1\langle g|-|r\rangle_1\langle r|)\right]\otimes \mathbb{I}\nonumber\\
H_2&=&\mathbb{I}\otimes f(wt)\left[\frac{\Omega_{0r}(t)}{2} e^{i\varphi_0(t)}| g \rangle_2 \langle r |+ \frac{\Omega_{1r}(t)}{2} e^{i\varphi_1(t)}| e \rangle_2 \langle r |+ {\rm H.c.}\right]\nonumber\\
&+&V|rr\rangle\langle rr|\nonumber\\
H_3&=&H_1,
\end{eqnarray}
which $\mathcal{F}=1$ and $\Delta_1=0$~[$\mathcal{F}=f_1$ and $\Delta_1=\Delta_1(t)$] for the DG-FGQC~(FGQC-FGQC) gates.

\subsection{Performances of CNOT and CT gates}\label{sec:4A}
For constructing a CNOT gate in three steps: (i)~We first apply a laser pulse to the control qubit with $\Omega_1 = 2 \times 2\pi~{\rm MHz}$ for a duration $T_1 = 0.25~{\rm \mu s}$. (ii)~Considering the target qubit, set $\theta = -\pi/2$ and $e^{-iC\tau} = -1$ with $C\tau = -\pi$, $T_2 = \tau$ being the duration of the second step, and $\omega \tau = k\pi,~\forall k\in \mathbb{N}^+$. Under the requirement of FGQC, we choose a set of feasible parameters
$\Omega_0=\Omega_1$, $V=20\Omega_0$, $k = 8$, $T_2 = \tau = 7.715~{\rm \mu s}$, 
$w \approx  3.2576~{\rm MHz}$, and $N \approx 0.5806~{\rm MHz}$. (iii)~Finally apply a laser pulse with $\Omega_1 = -2 \times 2\pi~{\rm MHz}$ to the control qubit for a duration of $T_3 = 0.25~{\rm \mu s}$.
	
In order to measure the performance of the gate better, we need to calculate the average fidelity. This involves calculating 
the fidelity for all possible initial states under a parameter variation such as the Rabi error $\delta$, summing these fidelities, 
and then dividing by the total number of initial states to obtain the fidelity image. This ensures that the gate parameters are 
applicable to all initial states. However, due to the infinite number of initial states, it is not convenient to calculate, so we employ a method using a finite number 
of initial density matrices to calculate fidelity, 
%We utilize $4^n$ initial density matrices to determine the average fidelity, where n is the number of qubits.
 considering the average gate fidelity~\cite{nielsen2002simple}
\begin{equation}\label{eq.18}
F(U,\varepsilon)= \frac{{\textstyle \sum_{j}}{\rm tr}\left[UU_j^\dagger U^\dagger \varepsilon
	(U_j) \right]+d^2}{d^2(d+1)},
\end{equation}
where $U$ represents the ideal logic gate, $U_j$ represents the tensor of 
Pauli matrices $\mathbb{I}$, $\sigma_x$, $\sigma_y$, $\sigma_z$ for single-qubit gate. Similarly, $U_j$ denotes $\mathbb{I}\otimes\mathbb{I}$, 
$\mathbb{I}\otimes \sigma_x$, $\mathbb{I}\otimes \sigma_y$, $\mathbb{I}\otimes \sigma_z$, $\cdots$, $\sigma_z\otimes \sigma_z$ for a 
two-qubit gate. Here $d = 4$ for the two-qubit gate, and $\varepsilon(U_j)$ denotes trace-preserving quantum operation. 

We set the relative Rabi error $\delta$ as a variable with $\delta \in \left [ -0.1, 0.1 \right ] $ to calculate the average fidelity, making the Rabi frequency deviations $\Omega_{0,1}(t)\rightarrow(1+\delta)\Omega_{0,1}(t)$. 
The DG-FGQC scheme, FGQC-FGQC scheme, and the original FGQC scheme all involve applications of Floquet theory on a two-qubit system. However, different from the original FGQC scheme shown in ~\ref{eqB}, where the laser is applied to two qubits under the control of parameter $\phi$ and the sum of squared laser frequencies on both qubits remains constant, our schemes ensure the sum of squared Rabi frequencies between the ground and excited states of the target qubit remains constant. We are uncertain which between the DG-FGQC and FGQC-FGQC schemes is more advantageous for the results, and whether our schemes maintain the high robustness against control errors seen in the original scheme. To address this, we compare the results of both DG-FGQC and FGQC-FGQC schemes against the original FGQC scheme. This comparison aims not only to validate the correctness and value of the FGQC schemes in our model but also to identify the scheme that exhibits superior performance.
The resulting average fidelity for the DG-FGQC CNOT gate is shown by the curve labeled Rydberg-blockade DG-FGQC CNOT in figure~\ref{fig2}.
Additionally, for the FGQC-FGQC scheme, we follow $\Delta_1(t)=\sin[\lambda_2'(t)]\Omega'$ and 
$\Omega_1(t)=\cos[\lambda_2'(t)]\Omega'$, where $\lambda_2'(t)=N_1t$~\cite{wang2021error}. And specify 
$\Omega'=2 \times 2\pi$~MHz, $\omega_1\approx 0.5133\times 2\pi$~MHz,
$N_1\approx45.728\times 2\pi$~kHz, $\tau_1\approx7.797~{\rm \mu s}$, and $\varphi_c=\pi/2$. Accordingly, we have $T_3=T_1=\tau_1$, and the 
resulting average fidelity plot for the FGQC-FGQC CNOT gate is shown by the curve labeled Rydberg-blockade FGQC-FGQC CNOT in figure~\ref{fig2}. At 
the same time, as shown in ~\ref{eqB}, the original two-qubit FGQC scheme only provides a SWAP-like gate for two-qubit and does not construct two-qubit controlled-X gates, etc. Therefore, we only compared it with our CNOT gate and CT gate to ensure the basic correctness of our scheme. So for reference we restored the fidelity of the two-qubit SWAP-like gate implemented with the 
original FGQC scheme~\cite{wang2021error},  as shown by the curve labeled original FGQC SWAP in figure~\ref{fig2}. By comparing the average fidelity curves under three different schemes, we find that applying the 
Floquet scheme in the Rydberg blockade regime performs better than the conventional approach. It exhibits higher fidelity and better robustness against the laser Rabi error, especially when both control and target qubits are operated with Floquet engineering. In particular, in the fidelity plot, the FG-FGQC and FGQC-FGQC gates do not achieve fidelity equal to 1 when $\delta=0$. The insignificant infidelity originates from the
Floquet adiabatic approximation.

\begin{figure}[htp]\centering
	\includegraphics[width=0.6\linewidth]{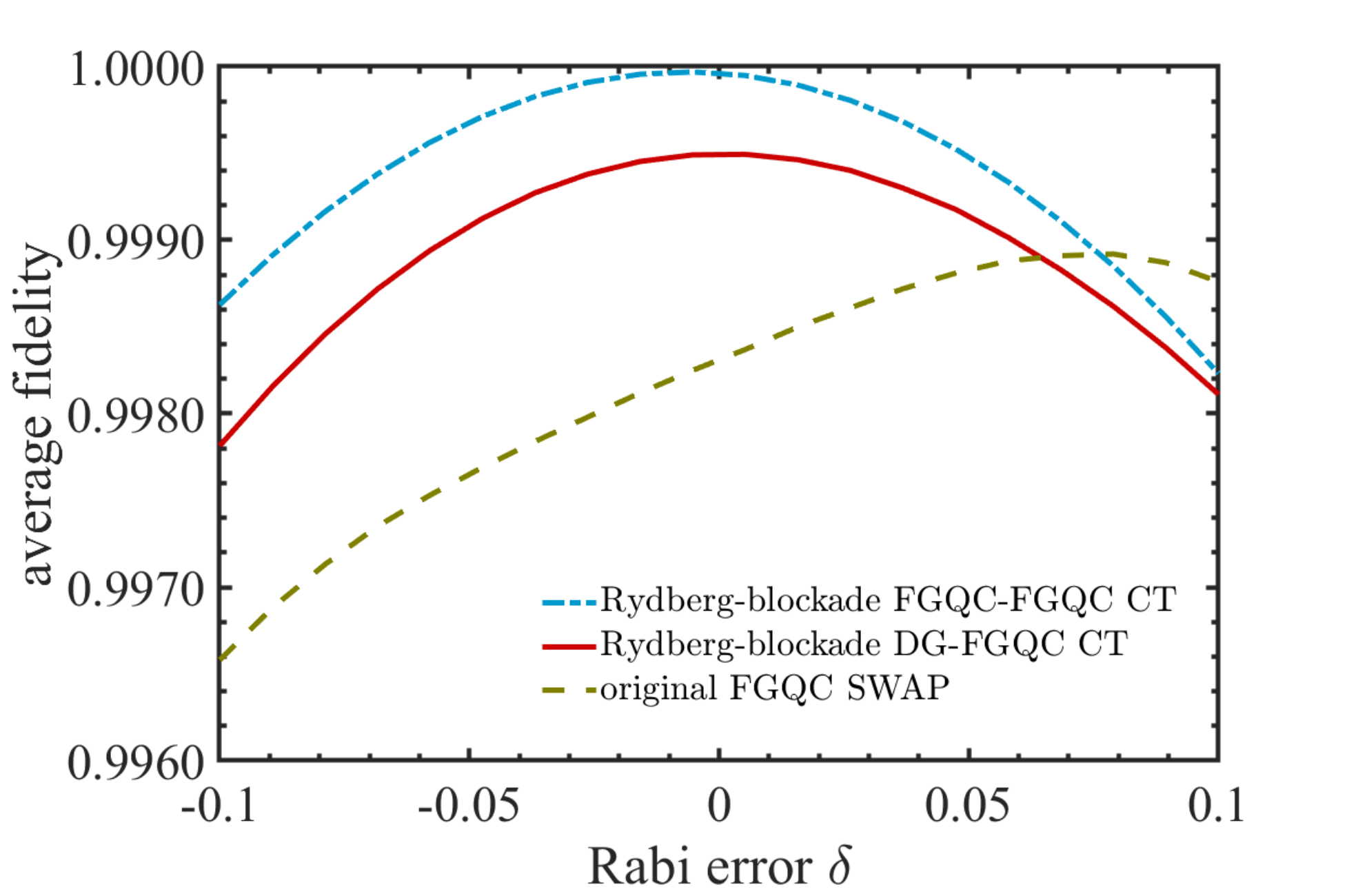}\\ 
	\caption{Comparison of the average fidelity for implementing a CT gate under our
	two new Floquet schemes, and implementing a SWAP-like gate under original Floquet
	scheme.}\label{fig3}
\end{figure}

For constructing a CT gate, the steps (i) to (iii) are similar to those for a CNOT gate. In the step (ii), considering the control qubit, we set $\theta=0$ and $e^{-iC\tau}=e^{i\pi/4}$. Due to $C = -\left(N/2\right)\left [1-J_{0}(\Omega_0/w)  \right ]$ and $\omega \tau = k\pi$, $\forall k\in \mathbb{N}^+$, we
choose a set of feasible parameters with $k=2$, $T_2=\tau=1.157~{\rm \mu s}$, $\omega \approx 5.41~{\rm MHz}$, 
and $N \approx 1.4191~{\rm MHz}$. There are still $\Omega_0=\Omega_1$ and $V=20\Omega_0$. In the same way, 
we set the relative Rabi error $\delta$ to change the Rabi frequencies  $\Omega_{0,1}(t)\rightarrow(1+\delta)\Omega_{0,1}(t)$. 
The resulting average fidelities for implementing the DG-FGQC and FGQC-FGQC CT gates are shown in figure~\ref{fig3}. Similarly, as a reference, the original FGQC SWAP-like gate is shown by the curve labeled original FGQC SWAP in figure ~\ref{fig3}.
The results indicate that, similar to the CNOT gate, the Floquet-engineering geometric CT gate can be implemented with high 
fidelity and excellent robustness against Rabi errors. Even when the relative Rabi error reaches $|\delta|=0.1$, the CT gate 
average fidelity is near 0.998. However, since the longer evolution time is required for implementing entangling gates by the 
Floquet engineering, it may not necessarily hold better performances when considering decay of atoms for the Floquet geometric 
gates. We will analyze this aspect in section ~\ref{sec:4B}.
\begin{figure}[htp]\centering
	\includegraphics[width=0.6\linewidth]{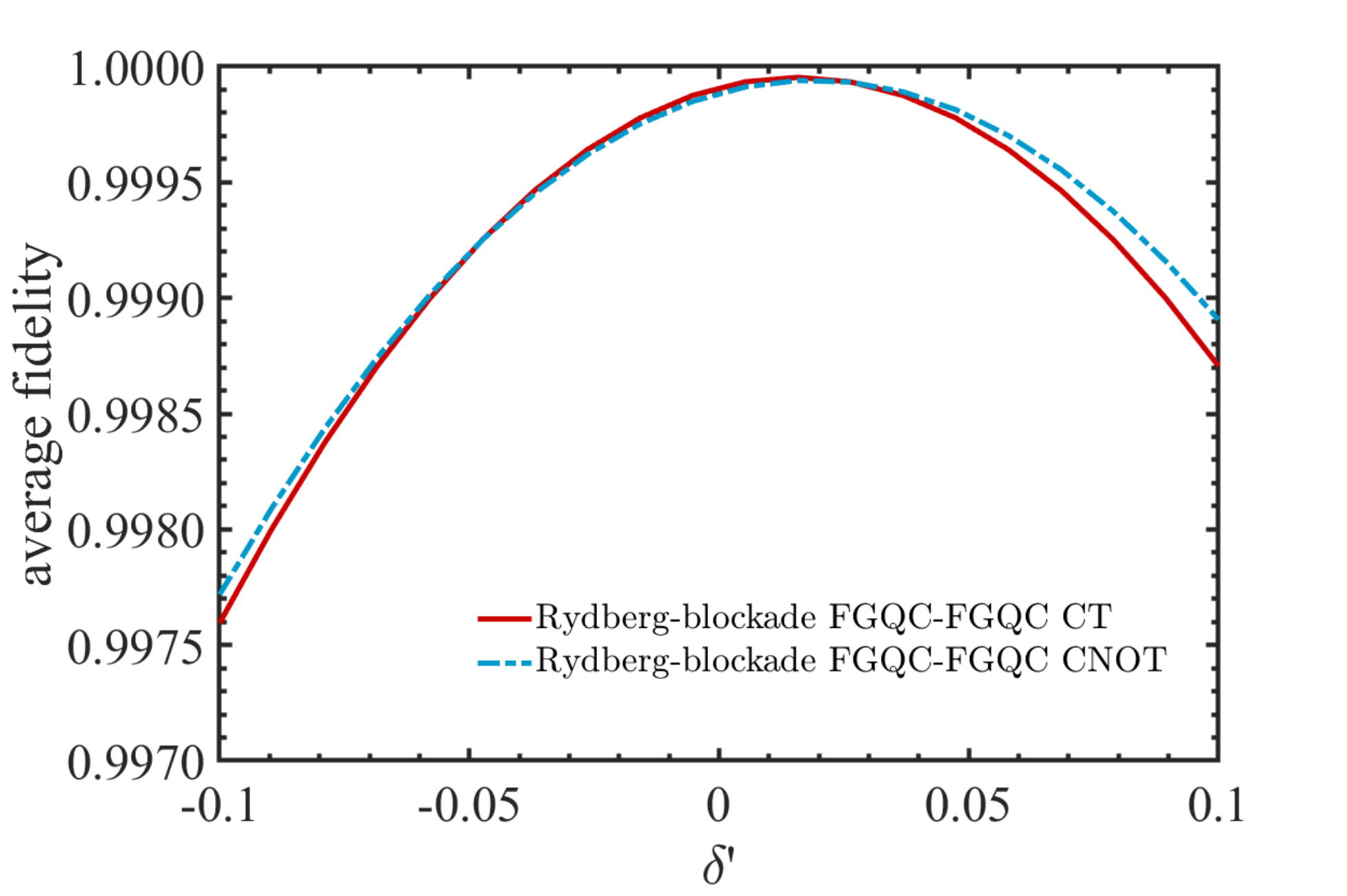}\\ 
	\caption{The impact of $\Delta_1(t)$ on two different gates under the FGQC-FGQC scheme.}\label{fig4}
\end{figure}

In addition to Rabi error $\delta$, we further analyze the impact of $\Delta_1(t)$ on gate fidelity when applying the FGQC-FGQC scheme. 
We set a relative error $\delta'$ to change the frequency detuning $\Delta_{1}(t)\rightarrow(1+\delta')\Delta_{1}(t)$ with $\delta' \in \left [ -0.1, 0.1 \right ] $. 
Then we get the resulting average fidelities for
implementing the FGQC-FGQC CNOT and FGQC-FGQC CT gates as shown in figure~\ref{fig4}. The fidelity plot demonstrates 
that the FGQC-FGQC scheme holds high fidelity and robustness when dealing with control qubit detuning error $\delta_1(t)$. 

\begin{figure}[htp]\centering
	\includegraphics[width=0.6\linewidth]{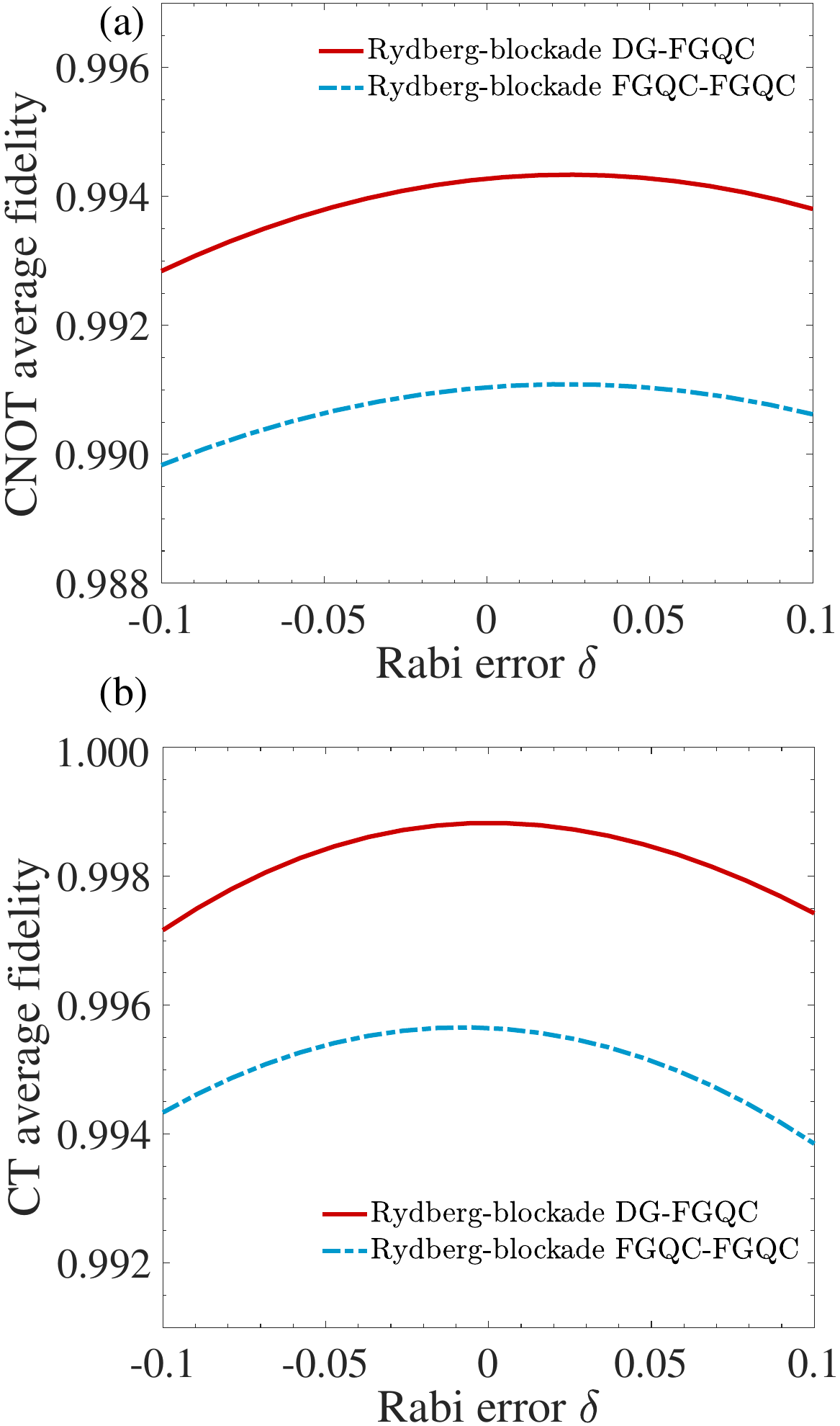}\\ 
	\caption{Impact of spontaneous emission on two different Floquet schemes for implementing (a)~CNOT and (b)~CT gates.}\label{fig5}
\end{figure}

\subsection{Floquet geometric entangling gate performance with atomic decay}\label{sec:4B}
We consider the dissipative dynamics of the system described by the following effective master equation~\cite{rao2013dark}:
\begin{eqnarray}\label{eq.19}
\frac{\partial \rho }{\partial t}&=&-i\left [ \mathcal{H}_I\rho -\rho \mathcal{H}_I^\dagger  \right ]
+\sum_{j=1}^{2}( \sigma _0^j\rho \sigma _0^{j\dagger } +
\sigma _1^j\rho \sigma _1^{j\dagger }) ,\nonumber\\
\mathcal{H}_I&=&H_I-\frac{i}{2}\sum_{j=1}^{2}( \sigma _0^j \sigma _0^{j\dagger } +
\sigma _1^j \sigma _1^{j\dagger }) ,    
\end{eqnarray}
in which $\rho$ is the density matrix of the system, and $H_I$ represents the system Hamiltonian given by equation~(\ref{three_step}). 

Here we specify $^{87}$Rb atoms as the qubit candidate, and the choose the atomic ground states $|g(e)\rangle=|5S_{1/2},F=1(2),m_F=0\rangle$~\cite{PhysRevA.82.030306}. The Rydberg state is chosen as $|r\rangle=|97S,J=1/2,m_J=1/2\rangle$~\cite{PhysRevLett.112.243601} with the spontaneous emission rate being $\Gamma = 2.48~{\rm kHz}$ at 300~K~\cite{PhysRevA.79.052504}. $\sigma _0^j=\sqrt{4\Gamma/13}\left | 0  \right \rangle_j \langle 1|$ and 
$\sigma _1^j=\sqrt{3\Gamma /13}  \left | 1  \right \rangle_j \langle 1|$ describe the decay processes of 
the $j$th atom by effective spontaneous emission~\cite{sun2021one}. We analyze the impact of spontaneous emission and 
compare two different schemes for implementing the CNOT and CT gates, and the obtained results are shown 
in figure~\ref{fig5}.

By comparing figure~\ref{fig5} and figures~\ref{fig2} and \ref{fig3}, it is evident that, although the fidelity and 
robustness to Rabi errors of the FGQC-FGQC scheme surpass the DG-FGQC scheme in the absence of dissipation, 
while when considering spontaneous emission, the fidelity of the FGQC-FGQC gate is lower that of the DG-FGQC gate. 
It is attributed to the high demands on the evolution time for FGQC, especially for CNOT gates, resulting in a significant impact of dissipation. In contrast, the DG-FGQC scheme only employs FGQC in the intermediate second step, which has a relatively minor impact. Therefore, it exhibits a preference for dissipation.
To further investigate the impact of dissipation, we selected gates constructed using the superior DG-FGQC scheme for a comprehensive scan analysis under different decay rate.

\begin{figure}[htp]\centering
	\includegraphics[width=0.6\linewidth]{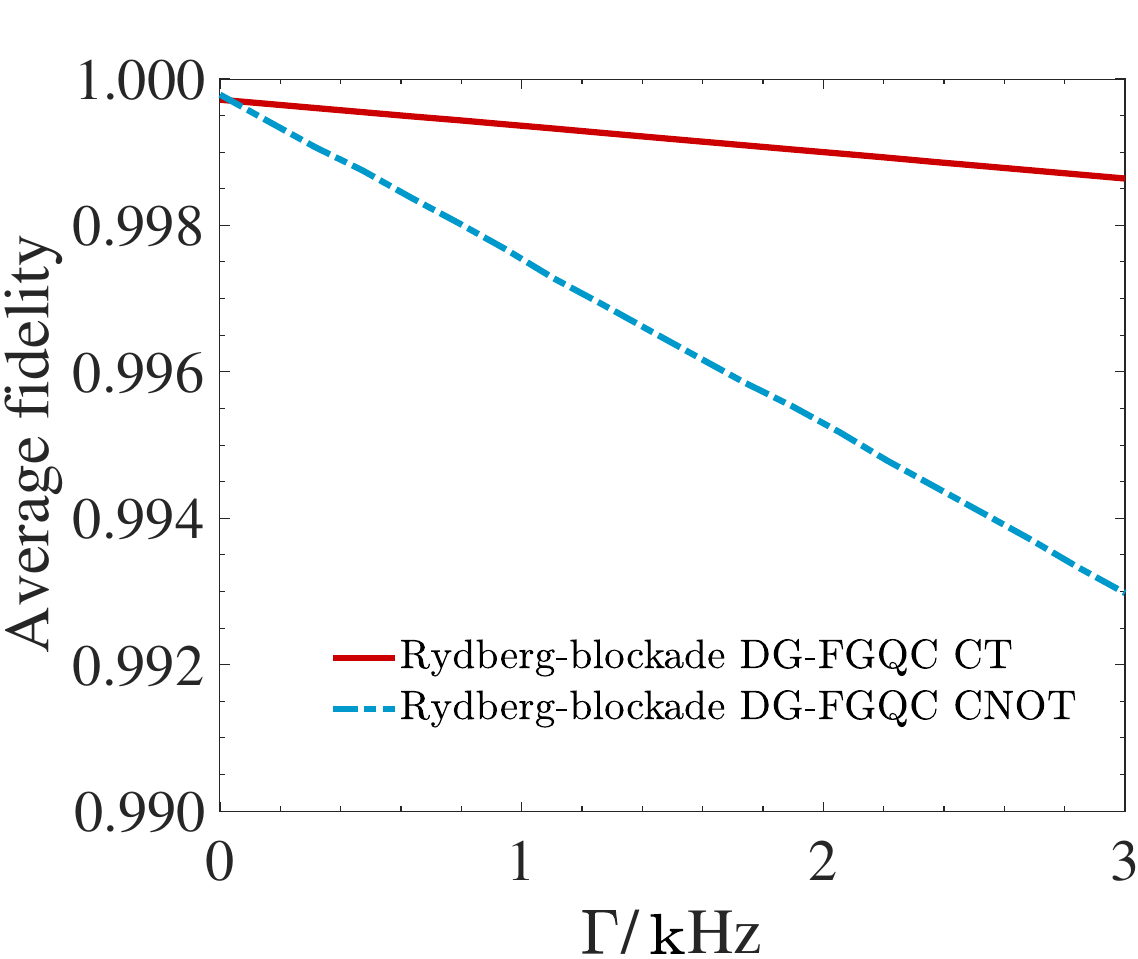}\\ 
	\caption{Effect of different atomic decay on fidelities of the DG-FGQC CT and CNOT gates.}\label{fig6}
\end{figure}
Finally we consider effect of different decay rates of atoms on the fidelity of implementing Floquet entangling gates, corresponding to different situations with different atomic temperatures in experiment. In figure~\ref{fig6} we obtain the average fidelity of implementing DG-FGQC CNOT and CT gates with different atomic decay rates. Then we can get that DG-FGQC CT gate exhibits a certain resistance to atomic decay compared to the CNOT gate, which is attributed to the shorter Floquet evolution time of the CT gate compared to the CNOT gate, being only about 15$\%$ of the 
latter. As a matter of fact, overall the DG-FGQC entangling gates are robust against atomic decay, because even when the decay rate reaches $\Gamma=3$~kHz the average fidelity of DG-FGQC entangling gates can be over 0.992.

\section{Conclusion}\label{sec:5}
In conclusion, we propose two-qubit entangling gate schemes with varying degrees of applying Floquet theory in the Rydberg-blockade regime. Utilizing two three-level Rydberg atoms and employing time-dependent or time-independent driving fields acting stepwise on different atoms, numerical simulations demonstrate that our approach exhibits higher robustness to control errors and higher fidelity in addressing system control errors compared to traditional methods. Through simulations considering spontaneous emission, we find that the DG-FGQC scheme outperforms the 
FGQC-FGQC scheme in terms of robustness, particularly for phase gates with shorter evolution times. Additionally, compared to reference Floquet schemes encoding computational state on Rydberg states, 
our approach encodes qubits on the two ground states of atoms, demonstrating superior 
decoherence resistance and lower requirements on experimental conditions and control precision. 
Therefore, our scheme shows feasibility in the application of Floquet theory.	
	
	\section*{Acknowledgements}
	We acknowledge supports from National Natural Science Foundation of China (12304407, 12274376) and China Postdoctoral Science Foundation (2023TQ0310, GZC20232446).
	\appendix
	\renewcommand{\appendixname}{Appendix}
	\section{{Offsetting the dynamical phase using a part of the geometric phase}}\label{eqA}
	For a system with a Hamiltonian $H\left(t\right)$, at $t=0$ we have a complete set of basis vectors 
	$\{|\psi_\alpha\left(0\right)\rangle\}$, and the time-dependent state $|\psi_\alpha\left(t\right)\rangle$ 
	satisfying the Schrödinger equation is given by $|\psi_\alpha\left(t\right)\rangle=U\left(t,0\right)|\psi_\alpha\left(0\right)\rangle$, 
	where  $U\left(t,0\right)$ denotes the time evolution operator from the initial time to time $t$. 
	We choose a new set of time-dependent basis $\{|\mu_\alpha\left(t\right)\rangle\}$ that satisfy the boundary 
	condition: $|\mu_\alpha\left(\tau\right)\rangle=|\mu_\alpha\left(0\right)\rangle=|\psi_\alpha\left(0\right)\rangle$. 
	The time-dependent state can be written as $|\psi_\alpha\left(t\right)\rangle=\sum_\beta c_{\alpha\beta}(t)|\mu_\beta\left(t\right)\rangle$. 
	By substituting it into the Schrödinger equation, we can obtain
	\begin{eqnarray}\label{eq.A1}
		\frac{\mathrm{d} }{\mathrm{d} x}
 		c_{\alpha \beta }\left ( t \right ) =i\sum_{\gamma}^{ } 
		\left [ A_{\alpha \gamma }\left ( t \right ) -H_{\alpha \gamma }\left ( t \right )  \right ]
		c_{\gamma \beta } \left ( t \right ) ,   	 
	\end{eqnarray}
	where $H_{\alpha  \beta }(t) =\left\langle\mu_\alpha(t) \right| H(t) \left | \mu_\beta(t)\right \rangle$, and
	\begin{eqnarray}\label{eq.A2}
		A_{\alpha  \beta }(t) =\left\langle\mu_\alpha(t) \right| i\partial_t \left | \mu_\beta(t)\right \rangle.  	 
	\end{eqnarray} 
	 
	Consider any unitary once differentiable operator $V(t)$ satisfying $V(\tau)=V(0)$. The operator $A(t)$ 
	is transformed as a proper gauge potential under the basis change, and $A(t)$ will generally contribute a 
	non-Abelian geometric phase to the temporal evolution operator $U(\tau, 0)$.

	Using a portion of geometric phase to offset dynamic phase~\cite{liu2020brachistochrone}, 
	assuming that the auxiliary state ${\mu_\alpha(t)}$ depends on two time-dependent parameters $\lambda_1(t)$ and $\lambda_2(t)$, yielding the gauge potential $A_{\alpha  \beta }(t)=A_{\alpha  \beta }^{(\lambda_1)}(t)+A_{\alpha  \beta }^{(\lambda_2)}(t)$. 
	Then, from equation~(\ref{eq.A2}), we can obtain
	\begin{eqnarray}\label{eq.A3}
		A_{\alpha  \beta }^{\lambda_j}(t) =\dot{\lambda_j}(t) \left\langle\mu_\alpha(\lambda) \right| i\frac{\partial}{\partial \lambda _j}  \left | \mu_\beta(\lambda)\right \rangle.  	 
	\end{eqnarray} 
	If $A_{\alpha  \beta }^{(\lambda_1)}(t)=H_{\alpha  \beta }(t)$, then the timeevolution operator is governed by a part of gauge potential
	$A_{\alpha  \beta }^{(\lambda_2)}(t)=A_{\alpha  \beta }(t)-H_{\alpha  \beta }(t)$. 
	Set the time evolution operator $R(\lambda)=e^{-iF[\lambda_1(t)]H_0[\lambda_2(t)]}$ for $\{\mu_\alpha(t)\}$, and then we can obtain a 
	new representation for $A_{\alpha  \beta }^{\lambda_1}(t)$ of equation~(\ref{eq.A2}):
	\begin{eqnarray}\label{eq.A4}
		A_{\alpha  \beta }^{\lambda_1}(t) =\left\langle\mu_\alpha(\lambda) \right|\left [i\dot{\lambda_1}(t) \frac{\partial (-i\mathcal{F} [\lambda_1(t)]H_0[\lambda_2(t)])}{\partial \lambda _1}\right ]\left | \mu_\beta(\lambda)\right \rangle.  	 
	\end{eqnarray}
	Therefore, we choose $H(t)=i\dot{\lambda_1}(t) \frac{\partial (-i\mathcal{F} [\lambda_1(t)]H_0[\lambda_2(t)])}{\partial \lambda _1}=\dot{\lambda_1}(t) \frac{\partial {\mathcal{F}}(\lambda_1)}{\partial \lambda _1} H_0(\lambda_2)$, and then we can obtain
	a pure geometric effective Hamiltonian 
	\begin{eqnarray}\label{eq.A5}
		H_{eff}(t)=A^{\lambda_2}(t)=\dot{\lambda_2}(t) R^{\dagger}(\lambda)i\partial_{\lambda_2} R(\lambda).
	\end{eqnarray}
	\section{{Original two-qubit FGQC scheme}}\label{eqB}
	Consider two two-level Rydberg atoms with RRI V and Rabi frequency $\Omega_\alpha(t)$. In the rotating 
	frame, the Hamiltonian of the two-Rydberg-atom system \cite{zhao2017rydberg} 
	$H_{12}=H_1\otimes \mathbb{I}\nonumber_2+\mathbb{I}\nonumber_1\otimes H_2+V\left | 11  \right \rangle \left \langle 11 \right |$,
	where $H_\alpha=\Omega_\alpha(t)(\left | 0  \right \rangle_\alpha \left\langle 1\right| +\left | 1  \right \rangle_\alpha \left\langle 0\right|)$
	is a single-atom Hamiltonian describing the interaction between the $\alpha$th atom and laser pulses; $\mathbb{I}\nonumber_\alpha$ 
	is the identity operator acting on the $\alpha$th Rydberg atom. Choose $\Omega _1(t)=-\Omega _R(t)\cos(\phi/2)$ and 
	$\Omega _2(t)=\Omega _R(t)\sin(\phi/2)$, then
	\begin{eqnarray}\label{eq.B1}
		H_{12}=\Omega_R(t)(\left | B  \right \rangle \left \langle 00 \right |-\left | B' \right \rangle \left \langle 11 \right |+\rm H.c.)+V\left | 11  \right \rangle \left \langle 11 \right |,  
	\end{eqnarray}
	where $\left|B\right\rangle =\sin(\phi/2)\left|01\right\rangle-\cos(\phi/2)\left|10\right\rangle$, $\left|B'\right\rangle =\cos(\phi/2)\left|01\right\rangle-\sin(\phi/2)\left|10\right\rangle$. 
	Take a rotation $U=\exp[-iVt\left | 11  \right \rangle \left \langle 11 \right |]$, then
	\begin{eqnarray}\label{eq.B2}
		H_{rot}(t)=\Omega_R(t)(\left | B  \right \rangle \left \langle 00 \right |-\left | B' \right \rangle \left \langle 11 \right |e^{-iVt})+\rm H.c. 
	\end{eqnarray}
	When $V\gg \Omega_R(t)$, the off-resonant terms are negligible, and the simultaneous excitation of two atoms 
	from the ground state to the Rydberg states is inhibited, thus
	\begin{eqnarray}\label{eq.B3}
		H_{rot}(t)\approx \Omega_R(t)\left | B  \right \rangle \left \langle 00 \right |+\rm {H.c.} 
	\end{eqnarray}
	Give the subspace $S=Span\{\left | B  \right \rangle, \left | 00  \right \rangle\}$ and $\Omega_R(t)=[\Omega_0/2]f(\omega t)\exp[i\varphi(t)]$, 
	then the Hamiltonian~(\ref{eq.B4}) can be rewritten as
	\begin{eqnarray}\label{eq.B4}
		H_{rot}(\omega t, t)= f(\omega t)\mathbf{F} \cdot \mathbf{r'}(t), 
	\end{eqnarray}
	where $\mathbf{F}\equiv \left(\tilde{\sigma}_x,\tilde{\sigma}_y,\tilde{\sigma}_z\right)^T/2$ and $\mathbf{r'}(t)=\Omega_0\left(\cos\varphi_1(t),-\sin\varphi_1(t),0\right)^T$ 
	with $\tilde{\sigma}_x=\left | B  \right \rangle \left\langle 00\right| +\left | 00  \right \rangle \left\langle B\right|$, 
	$\tilde{\sigma}_y=-i\left | B  \right \rangle \left\langle 00\right| +i\left | 00  \right \rangle \left\langle B\right|$, 
	and $\tilde{\sigma}_z=\left | B  \right \rangle \left\langle B\right| -\left | 00  \right \rangle \left\langle 00\right|$ 
	Consider $f(wt)=\cos(wt)$, then $R(t)=\exp[-i\sin(wt)\mathbf{F}\cdot \mathbf{r'}(t)/w]$, satisfying the boundary condition $R(\tau)=R(0)=I$ if $\omega\tau=k\pi$, 
	$\forall k\in N^+$. Similar to obtaining equation (\ref{eq.6}), we obtain
	\begin{eqnarray}\label{eq.B5}
		H_t(t) =  \dot{\varphi}(t)R^{\dagger}(t)(-i\partial /\partial\varphi)R(t).   
	\end{eqnarray}
	If $\omega\gg\dot{\varphi}(t)$, by referring to equation (39) in reference~\cite{novivcenko2019non}, and for the same reason as obtaining equation~(\ref{eq.7}), the effective Hamiltonian can be obtained as
    \begin{eqnarray}\label{eq.B6}
		H_{\rm eff}^{(0)}(t)=[1-J_{0}(\Omega_0/w)]\mathbf{F}\cdot(\mathbf{r}\times \dot{\mathbf{r'}})/\Omega_0^2.    
	\end{eqnarray}
	If $\varphi(t)=Nt$, 
	then $H_{eff}^{(0)}(t)=-N[1-J_{0}(\Omega_0/w)]\tilde{\sigma}_z/2$. In the basis $\{\left | 00  \right \rangle, \left | 01  \right \rangle, \left | 10  \right \rangle, \left | 11  \right \rangle\}$, 
	the matrix of the effective Hamiltonian can be written as 
	\begin{eqnarray}\label{eq.B7}
		H_{eff}^{(0)}(t)=C\cdot\left(
			\begin{array}{cccc}
			0&0&0&0\\
			0&\cos^2(\frac{\phi}{2})&-\sin(\frac{\phi}{2})&0\\
			0&-\sin(\frac{\phi}{2}) &\sin^2(\frac{\phi}{2})&0\\
			0&0&0&-1
			\end{array}
		\right),  
	\end{eqnarray}   
	where $C\equiv$$-\left(N/2\right)\left [1-J_{0}(\Omega_0/w)  \right ] $. Then, through 
	$U=\mathcal{T}e^{-i\int_{0}^{\tau} H_{\rm eff}^{(0)}(t)dt}$, the time-evolution operator is obtained
	\begin{eqnarray}\label{eq.B8}
		U(\tau, 0)=\left(
			\begin{array}{cccc}
			1&0&0&0\\
			0&\frac{1}{2}(D_1-D_2\cos\phi)&\frac{1}{2}D_2\sin\phi&0\\
			0&\frac{1}{2}D_2\sin\phi&\frac{1}{2}(D_1+D_2\cos\phi)&0\\
			0&0&0&e^{iC\tau}
			\end{array}
		\right),  
	\end{eqnarray} 
	where $D_1=1+e^{-iC\tau}$, $D_2=1-e^{-iC\tau}$. $e^{iC\tau}=-1$ and $\phi=\pi/2$ yield a SWAP-like gate
	\begin{eqnarray}\label{eq.B9}
		U(\tau, 0)=\left(
			\begin{array}{cccc}
			1&0&0&0\\
			0&0&1&0\\
			0&1&0&0\\
			0&0&0&-1
			\end{array}
		\right).  
	\end{eqnarray} 
	\section*{References}
	\bibliographystyle{iopart-num}
	\bibliography{iopart-num}

\providecommand{\newblock}{}
\begin{thebibliography}{10}
\expandafter\ifx\csname url\endcsname\relax
  \def\url#1{{\tt #1}}\fi
\expandafter\ifx\csname urlprefix\endcsname\relax\def\urlprefix{URL }\fi
\providecommand{\eprint}[2][]{\url{#2}}

\bibitem{berry1984quantal}
Berry M~V 1984 {\em Proceedings of the Royal Society of London. A. Mathematical and Physical Sciences\/} {\bf 392} 45--57 \urlprefix\url{https://royalsocietypublishing.org/doi/10.1098/rspa.1984.0023}

\bibitem{aharonov1987phase}
Aharonov Y and Anandan J 1987 {\em Physical Review Letters\/} {\bf 58} 1593 \urlprefix\url{https://doi.org/10.1103/PhysRevLett.58.1593}

\bibitem{wilczek1984appearance}
Wilczek F and Zee A 1984 {\em Physical Review Letters\/} {\bf 52} 2111 \urlprefix\url{https://doi.org/10.1103/PhysRevLett.52.2111}

\bibitem{anandan1988non}
Anandan J 1988 {\em Physics Letters A\/} {\bf 133} 171--175 \urlprefix\url{https://linkinghub.elsevier.com/retrieve/pii/0375960188910109}

\bibitem{vandersypen2001experimental}
Vandersypen L~M, Steffen M, Breyta G, Yannoni C~S, Sherwood M~H and Chuang I~L 2001 {\em Nature\/} {\bf 414} 883--887 \urlprefix\url{https://www.nature.com/articles/414883a}

\bibitem{xu2012quantum}
Xu N, Zhu J, Lu D, Zhou X, Peng X and Du J 2012 {\em Physical Review Letters\/} {\bf 108} 130501 \urlprefix\url{https://link.aps.org/doi/10.1103/PhysRevLett.108.130501}

\bibitem{martin2012experimental}
Martin-Lopez E, Laing A, Lawson T, Alvarez R, Zhou X~Q and O'brien J~L 2012 {\em Nature Photonics\/} {\bf 6} 773--776 \urlprefix\url{https://arxiv.org/abs/1111.4147}

\bibitem{rebentrost2014quantum}
Rebentrost P, Mohseni M and Lloyd S 2014 {\em Physical Review Letters\/} {\bf 113} 130503 \urlprefix\url{https://journals.aps.org/prl/abstract/10.1103/PhysRevLett.113.130503}

\bibitem{li2015experimental}
Li Z, Liu X, Xu N and Du J 2015 {\em Physical Review Letters\/} {\bf 114} 140504 \urlprefix\url{https://journals.aps.org/prl/abstract/10.1103/PhysRevLett.114.140504}

\bibitem{cong2019quantum}
Cong I, Choi S and Lukin M~D 2019 {\em Nature Physics\/} {\bf 15} 1273--1278 \urlprefix\url{https://arxiv.org/abs/1810.03787}

\bibitem{de2003berry}
De~Chiara G and Palma G~M 2003 {\em Physical Review Letters\/} {\bf 91} 090404 \urlprefix\url{https://link.aps.org/doi/10.1103/PhysRevLett.91.090404}

\bibitem{zhu2005geometric}
Zhu S~L and Zanardi P 2005 {\em Physical Review A\/} {\bf 72} 020301 \urlprefix\url{https://link.aps.org/doi/10.1103/PhysRevA.72.020301}

\bibitem{leek2007observation}
Leek P~J, Fink J, Blais A, Bianchetti R, Goppl M, Gambetta J~M, Schuster D~I, Frunzio L, Schoelkopf R~J and Wallraff A 2007 {\em Science\/} {\bf 318} 1889--1892 \urlprefix\url{https://arxiv.org/abs/0711.0218}

\bibitem{filipp2009experimental}
Filipp S, Klepp J, Hasegawa Y, Plonka-Spehr C, Schmidt U, Geltenbort P and Rauch H 2009 {\em Physical Review Letters\/} {\bf 102} 030404 \urlprefix\url{https://link.aps.org/doi/10.1103/PhysRevLett.102.030404}

\bibitem{berger2013exploring}
Berger S, Pechal M, Abdumalikov~Jr A~A, Eichler C, Steffen L, Fedorov A, Wallraff A and Filipp S 2013 {\em Physical Review A\/} {\bf 87} 060303 \urlprefix\url{https://link.aps.org/doi/10.1103/PhysRevA.87.060303}

\bibitem{jones2000geometric}
Jones J~A, Vedral V, Ekert A and Castagnoli G 2000 {\em Nature\/} {\bf 403} 869--871 \urlprefix\url{https://pubmed.ncbi.nlm.nih.gov/10706278/}

\bibitem{wu2005holonomic}
Wu L~A, Zanardi P and Lidar D 2005 {\em Physical Review Letters\/} {\bf 95} 130501 \urlprefix\url{https://link.aps.org/doi/10.1103/PhysRevLett.95.130501}

\bibitem{wu2013geometric}
Wu H, Gauger E~M, George R~E, M{\"o}tt{\"o}nen M, Riemann H, Abrosimov N~V, Becker P, Pohl H~J, Itoh K~M, Thewalt M~L {\em et~al.\/} 2013 {\em Physical Review A\/} {\bf 87} 032326 \urlprefix\url{https://link.aps.org/doi/10.1103/PhysRevA.87.032326}

\bibitem{huang2019experimental}
Huang Y~Y, Wu Y~K, Wang F, Hou P~Y, Wang W~B, Zhang W~G, Lian W~Q, Liu Y~Q, Wang H~Y, Zhang H~Y {\em et~al.\/} 2019 {\em Physical Review Letters\/} {\bf 122} 010503 \urlprefix\url{https://link.aps.org/doi/10.1103/PhysRevLett.122.010503}

\bibitem{zanardi1999holonomic}
Zanardi P and Rasetti M 1999 {\em Physics Letters A\/} {\bf 264} 94--99 \urlprefix\url{https://arxiv.org/abs/quant-ph/9904011}

\bibitem{duan2001geometric}
Duan L~M, Cirac J~I and Zoller P 2001 {\em Science\/} {\bf 292} 1695--1697 \urlprefix\url{https://arxiv.org/abs/quant-ph/0111086}

\bibitem{Wu_2019}
Wu J~L and Su S~L 2019 {\em Journal of Physics A: Mathematical and Theoretical\/} {\bf 52} 335301 \urlprefix\url{https://dx.doi.org/10.1088/1751-8121/ab2a92}

\bibitem{xiang2001nonadiabatic}
Xiang-Bin W and Keiji M 2001 {\em Physical Review Letters\/} {\bf 87} 097901 \urlprefix\url{https://link.aps.org/doi/10.1103/PhysRevLett.87.097901}

\bibitem{zhu2002implementation}
Zhu S~L and Wang Z 2002 {\em Physical Review Letters\/} {\bf 89} 097902 \urlprefix\url{https://link.aps.org/doi/10.1103/PhysRevLett.89.097902}

\bibitem{thomas2011robustness}
Thomas J, Lababidi M and Tian M 2011 {\em Physical Review A\/} {\bf 84} 042335 \urlprefix\url{https://link.aps.org/doi/10.1103/PhysRevA.84.042335}

\bibitem{zhao2017rydberg}
Zhao P, Cui X~D, Xu G, Sj{\"o}qvist E and Tong D 2017 {\em Physical Review A\/} {\bf 96} 052316 \urlprefix\url{https://link.aps.org/doi/10.1103/PhysRevA.96.052316}

\bibitem{li2020approach}
Li K, Zhao P and Tong D 2020 {\em Physical Review Research\/} {\bf 2} 023295 \urlprefix\url{https://link.aps.org/doi/10.1103/PhysRevResearch.2.023295}

\bibitem{chen2018nonadiabatic}
Chen T and Xue Z~Y 2018 {\em Physical Review Applied\/} {\bf 10} 054051 \urlprefix\url{https://link.aps.org/doi/10.1103/PhysRevApplied.10.054051}

\bibitem{zhang2020high}
Zhang C, Chen T, Li S, Wang X and Xue Z~Y 2020 {\em Physical Review A\/} {\bf 101} 052302 \urlprefix\url{https://link.aps.org/doi/10.1103/PhysRevA.101.052302}

\bibitem{liu2019plug}
Liu B~J, Song X~K, Xue Z~Y, Wang X and Yung M~H 2019 {\em Physical Review Letters\/} {\bf 123} 100501 \urlprefix\url{https://link.aps.org/doi/10.1103/PhysRevLett.123.100501}

\bibitem{sjoqvist2012non}
Sj{\"o}qvist E, Tong D~M, Andersson L~M, Hessmo B, Johansson M and Singh K 2012 {\em New Journal of Physics\/} {\bf 14} 103035 \urlprefix\url{https://iopscience.iop.org/article/10.1088/1367-2630/14/10/103035/meta}

\bibitem{xu2012nonadiabatic}
Xu G, Zhang J, Tong D, Sj{\"o}qvist E and Kwek L 2012 {\em Physical Review Letters\/} {\bf 109} 170501 \urlprefix\url{https://link.aps.org/doi/10.1103/PhysRevLett.109.170501}

\bibitem{xue2015universal}
Xue Z~Y, Zhou J and Wang Z 2015 {\em Physical Review A\/} {\bf 92} 022320 \urlprefix\url{https://link.aps.org/doi/10.1103/PhysRevA.92.022320}

\bibitem{xue2017nonadiabatic}
Xue Z~Y, Gu F~L, Hong Z~P, Yang Z~H, Zhang D~W, Hu Y and You J 2017 {\em Physical Review Applied\/} {\bf 7} 054022 \urlprefix\url{https://link.aps.org/doi/10.1103/PhysRevApplied.7.054022}

\bibitem{zhou2018fast}
Zhou J, Liu B, Hong Z and Xue Z 2018 {\em Science China Physics, Mechanics \& Astronomy\/} {\bf 61} 1--7 \urlprefix\url{https://link.springer.com/article/10.1007/s11433-017-9119-8}

\bibitem{hong2018implementing}
Hong Z~P, Liu B~J, Cai J~Q, Zhang X~D, Hu Y, Wang Z and Xue Z~Y 2018 {\em Physical Review A\/} {\bf 97} 022332 \urlprefix\url{https://link.aps.org/doi/10.1103/PhysRevA.97.022332}

\bibitem{mousolou2017electric}
Mousolou V~A 2017 {\em Physical Review A\/} {\bf 96} 012307 \urlprefix\url{https://link.aps.org/doi/10.1103/PhysRevA.96.012307}

\bibitem{zhao2020general}
Zhao P, Li K, Xu G and Tong D 2020 {\em Physical Review A\/} {\bf 101} 062306 \urlprefix\url{https://link.aps.org/doi/10.1103/PhysRevA.101.062306}

\bibitem{johansson2012robustness}
Johansson M, Sj{\"o}qvist E, Andersson L~M, Ericsson M, Hessmo B, Singh K and Tong D 2012 {\em Physical Review A\/} {\bf 86} 062322 \urlprefix\url{https://link.aps.org/doi/10.1103/PhysRevA.86.062322}

\bibitem{zheng2016comparison}
Zheng S~B, Yang C~P and Nori F 2016 {\em Physical Review A\/} {\bf 93} 032313 \urlprefix\url{https://link.aps.org/accepted/10.1103/PhysRevA.93.032313}

\bibitem{ramberg2019environment}
Ramberg N and Sj{\"o}qvist E 2019 {\em Physical Review Letters\/} {\bf 122} 140501 \urlprefix\url{https://link.aps.org/doi/10.1103/PhysRevLett.122.140501}

\bibitem{jing2017non}
Jing J, Lam C~H and Wu L~A 2017 {\em Physical Review A\/} {\bf 95} 012334 \urlprefix\url{https://link.aps.org/doi/10.1103/PhysRevA.95.012334}

\bibitem{liu2021super}
Liu B~J, Wang Y~S and Yung M~H 2021 {\em Physical Review Research\/} {\bf 3} L032066 \urlprefix\url{https://arxiv.org/pdf/2008.02176}

\bibitem{novivcenko2017floquet}
Novi{\v{c}}enko V, Anisimovas E and Juzeli{\=u}nas G 2017 {\em Physical Review A\/} {\bf 95} 023615 \urlprefix\url{https://link.aps.org/doi/10.1103/PhysRevA.95.023615}

\bibitem{novivcenko2019non}
Novi{\v{c}}enko V and Juzeli{\=u}nas G 2019 {\em Physical Review A\/} {\bf 100} 012127 \urlprefix\url{https://link.aps.org/doi/10.1103/PhysRevA.100.012127}

\bibitem{bomantara2018quantum}
Bomantara R~W and Gong J 2018 {\em Physical Review B\/} {\bf 98} 165421 \urlprefix\url{https://link.aps.org/doi/10.1103/PhysRevB.98.165421}

\bibitem{bomantara2018simulation}
Bomantara R~W and Gong J 2018 {\em Physical Review Letters\/} {\bf 120} 230405 \urlprefix\url{https://link.aps.org/doi/10.1103/PhysRevLett.120.230405}

\bibitem{wang2021error}
Wang Y~S, Liu B~J, Su S~L and Yung M~H 2021 {\em Physical Review Research\/} {\bf 3} 033010 \urlprefix\url{https://link.aps.org/doi/10.1103/PhysRevResearch.3.033010}

\bibitem{cooke2024investigation}
Cooke L~W, Tashchilina A, Protter M, Lindon J, Ooi T, Marsiglio F, Maciejko J and LeBlanc L~J 2024 {\em Physical Review Research\/} {\bf 6} 013057 \urlprefix\url{https://journals.aps.org/prresearch/abstract/10.1103/PhysRevResearch.6.013057}

\bibitem{jaksch2000fast}
Jaksch D, Cirac J~I, Zoller P, Rolston S~L, C{\^o}t{\'e} R and Lukin M~D 2000 {\em Physical Review Letters\/} {\bf 85} 2208 \urlprefix\url{https://link.aps.org/doi/10.1103/PhysRevLett.85.2208}

\bibitem{saffman2010quantum}
Saffman M, Walker T~G and M{\o}lmer K 2010 {\em Reviews of Modern Physics\/} {\bf 82} 2313 \urlprefix\url{https://link.aps.org/doi/10.1103/RevModPhys.82.2313}

\bibitem{petrosyan2017high}
Petrosyan D, Motzoi F, Saffman M and M{\o}lmer K 2017 {\em Physical Review A\/} {\bf 96} 042306 \urlprefix\url{https://link.aps.org/doi/10.1103/PhysRevA.96.042306}

\bibitem{levine2018high}
Levine H, Keesling A, Omran A, Bernien H, Schwartz S, Zibrov A~S, Endres M, Greiner M, Vuleti{\'c} V and Lukin M~D 2018 {\em Physical Review Letters\/} {\bf 121} 123603 \urlprefix\url{https://arxiv.org/abs/1806.04682}

\bibitem{PhysRevApplied.16.064031}
Wu J~L, Wang Y, Han J~X, Jiang Y, Song J, Xia Y, Su S~L and Li W 2021 {\em Physical Review Applied\/} {\bf 16}(6) 064031 \urlprefix\url{https://link.aps.org/doi/10.1103/PhysRevApplied.16.064031}

\bibitem{sun2021one}
Sun L~N, Yan L~L, Su S~L and Jia Y 2021 {\em Physical Review Applied\/} {\bf 16} 064040 \urlprefix\url{https://link.aps.org/doi/10.1103/PhysRevApplied.16.064040}

\bibitem{nielsen2002simple}
Nielsen M~A 2002 {\em Physics Letters A\/} {\bf 303} 249--252 \urlprefix\url{https://arxiv.org/abs/quant-ph/0205035}

\bibitem{rao2013dark}
Rao D~B and M{\o}lmer K 2013 {\em Physical Review Letters\/} {\bf 111} 033606 \urlprefix\url{https://link.aps.org/doi/10.1103/PhysRevLett.111.033606}

\bibitem{PhysRevA.82.030306}
Zhang X~L, Isenhower L, Gill A~T, Walker T~G and Saffman M 2010 {\em Physical Review A\/} {\bf 82}(3) 030306 \urlprefix\url{https://link.aps.org/doi/10.1103/PhysRevA.82.030306}

\bibitem{PhysRevLett.112.243601}
Li W, Viscor D, Hofferberth S and Lesanovsky I 2014 {\em Physical Review Letters\/} {\bf 112}(24) 243601 \urlprefix\url{https://link.aps.org/doi/10.1103/PhysRevLett.112.243601}

\bibitem{PhysRevA.79.052504}
Beterov I~I, Ryabtsev I~I, Tretyakov D~B and Entin V~M 2009 {\em Physical Review A\/} {\bf 79}(5) 052504 \urlprefix\url{https://link.aps.org/doi/10.1103/PhysRevA.79.052504}

\bibitem{liu2020brachistochrone}
Liu B~J, Xue Z~Y and Yung M~H 2020 {\em arXiv preprint arXiv:2001.05182\/} \urlprefix\url{https://arxiv.org/abs/2001.05182}

\end{thebibliography}
\end{document}